# A Concept of LNA with Low Input Impedance Using Common Base BJT for Low-Field MRI

Aleksei A. Nasonov[1], Mikhail V. Murzin[1], Nikolay V. Anisimov[2], Vasily S. Severikov[1], Anna A. Dyatlovich[1], Anna A. Hurshkainen[1], Georgiy A. Solomakha[3]

[1]School of Physics and Engineering, ITMO University, Saint Petersburg, Russian Federation

[2]Faculty of Chemistry, Lomonosov Moscow State University, Moscow, Russian Federation

[3]Max Planck Institute for Biological Cybernetics, Tübingen, Germany

**Corresponding author:**
Aleksei Nasonov, School of Physics and Engineering, ITMO University,
Kronverksky Pr. 49, bldg. A, St. Petersburg, 197101, Russia.
Email: aleksei.nasonov@metalab.ifmo.ru

**Purpose**: In magnetic resonance imaging (MRI), preamplifier decoupling is used to improve the signal-to-noise ratio of a multichannel receive coil array. In low-field and ultra-low-field MRI, low-noise amplifier (LNA) solutions with low input impedance (LII) are poorly represented and consist mostly of field-effect transistors. We propose a low-cost LNA based on a common-base circuit with bipolar transistors to achieve LII (<5 Ω), a low noise figure (<1.3 dB), and high gain (>30 dB).

**Methods**: The proposed LNA circuits were simulated, fabricated, and tested at frequencies of 3 MHz and 21.2 MHz. The decoupling capabilities of the proposed LNA were experimentally evaluated, and MR imaging was performed in a 0.5T MRI scanner alongside a comparative study with a commercial LNA. At 3 MHz, the LNA was evaluated by analyzing free induction decay signals.

**Results:** Proposed LNA solutions demonstrate a low noise figure below 1.3 dB with an input impedance of 3 Ω. It was also shown that the presented LNA circuit enables high SNR values in MR imaging. The developed amplifiers demonstrate superior performance compared to available commercial options within the selected HF band.

**Conclusion**: Due to the availability and cost-effectiveness of the selected transistors, the proposed LNA circuit is highly suitable for low-field MRI applications.

**Keywords:** preamplifier, low noise, decoupling, low-field MRI

**Word count** (body text): 2719

## 1 INTRODUCTION

Multichannel arrays of receive elements, commonly known as RF coils, are widely used in modern MRI systems to obtain high-resolution images (1). Besides improving the image signal-to-noise ratio (SNR), the use of receive arrays enables accelerated data acquisition via various parallel imaging techniques (2–4). However, minimizing channel coupling is crucial for parallel imaging acquisition to prevent SNR degradation. To address this issue, various decoupling methods have been proposed over the past four decades. Since array channels are typically composed of surface loops, their coupling is primarily caused by mutual inductance. One of the widely used techniques for adjacent-element decoupling is geometric overlapping (1). To decouple both adjacent and non-adjacent elements, decoupling methods utilizing a transformer (5) and a capacitor (6) can be used. For decoupling distant coil elements, or when circuit-based decoupling methods are not applicable, preamplifier decoupling is widely used (1). However, preamplifier decoupling relies on the input impedance of the first-stage low-noise amplifier (LNA), which is connected directly to the array element output to maximize the MR signal SNR. To achieve effective decoupling, either a high (~1 kΩ) or a low (~1-5 Ohm, should be used in combination with some type of impedance transformer) input impedance can be utilized.

Several studies (7–9) describe various preamplifier circuits that not only provide high gain and low noise but also exhibit the low input impedance (LII) required for preamplifier decoupling. However, the majority of these studies are tailored to frequency ranges associated with high-field MRI systems (1.5T and above). According to the literature (10), the lowest operating frequency for an LII LNA is 32.1 MHz (the Larmor frequency of $^{13}$C at 3T). Notably, such preamplifiers are typically based on obsolete HEMTs, such as the Broadcom/Avago ATF series. Alternatively, an OEM LNA produced by WanTcom (11) can be employed; however, its lowest specified frequency is limited to 13 MHz. With the rising interest in low field MRI (0.05 to 0.5T) (12–15) there is a growing demand for low-cost LNAs operating within this frequency range (approximately 2.1 to 21.3 MHz), especially considering the intensive development of portable MR scanners for developing countries. When employing the preamplifier decoupling method for array design, either a low or a high preamplifier input impedance is strictly required.

In this work, the focus is on designing a low-noise, low-impedance MRI preamplifier utilizing commercially available, low-cost BFR193 (16) bipolar junction transistors (BJTs).

Given the widespread use of BFR193, a manufacturer-provided SPICE model is available for numerical simulations. This facilitates efficient numerical simulation and optimization of the LNA design parameters for any target frequency. To achieve a low input impedance, a common-base configuration was employed as the first stage of the MRI LNA. To achieve a higher gain (~30 dB), which is critical for MRI receiver chains, a second common-emitter amplification stage was added. The LNA performance was simulated at frequencies of 3 MHz (0.07T $^{1}$H MRI) and 21.2 MHz (0.5T $^{1}$H MRI). These frequencies were selected to highlight the versatility of the proposed LII LNA design across various low-field MRI operating frequencies. At 21.3 MHz (0.5T), the LNA performance was evaluated via bench measurements and verified during MR imaging experiments. For the 2.98 MHz (0.07T) setup, bench testing was performed to compare the proposed LII LNA with a 50 Ω LNA implemented in a common-emitter configuration utilizing the same BFR193 transistor, with free induction decay (FID) signal acquisition used for evaluation.

## 2 **METHODS**

### **2. 1 Numerical simulations**

Using the Infineon BJT component library, an LNA circuit was designed in ADS 2024 (Keysight Technologies). DC simulation was used to obtain operating-point data, S-parameter simulation to obtain impedance characteristics, and a harmonic balance analysis to determine the compression point. The simulation model consisted of two stages, with a common base (CB) in the first stage and a common emitter (CE) in the second (Figure 1). Using the CB configuration as the first stage provided a low input impedance. However, since a single CB stage is limited to ~16 dB of power gain, a second CE configuration was introduced to boost the total LNA gain. If the order of the stages is reversed, the well-known cascode amplifier is obtained, offering all its typical advantages. However, our ‘inverted cascode’ circuit represents two independent stages, where each stage addresses a specific design challenge.

Capacitors $C_1$, $C_4$, and C6 are utilized for DC blocking. Resistors $R_1$ and $R_2$ form the biasing network for the first-stage transistor while simultaneously enhancing the amplifier's stability. Capacitor $C_2$ is connected in parallel with $R_1$ to shunt the resistor, grounding the base of $Q_1$ at high frequencies and enabling common-base operation. To provide a DC path to ground for the first stage, the emitter of transistor $Q_1$ is connected to ground via inductor L2. It should be noted that the parasitic resistance of the inductor affects the circuit's input impedance;

therefore an inductor with minimal series resistance it is recommended. Resistor $R_5$ is used both to adjust the output impedance and to improve amplifier stability by reducing the Q-factor of inductor $L_3$. In simulations, the parasitic resistance of $L_{1-4}$ was set to 1 Ω based on known values for commercial non-magnetic RF chokes (e.g., the Delevan 1840 series). To achieve minimum noise figure (NF) and high gain, the output impedance of the first CB stage must be properly matched to the input impedance of the second CE stage. This LNA circuit includes a two-element matching network consisting of a series capacitor, $C_t$, and a shunt inductor, $L_4$ – to compensate for the high output impedance of CB configuration. The CE stage was identical at both evaluated frequencies, as it is inherently broadband and does not require reactive matching elements. Input impedance of this stage is primarily determined by resistor $R_4$, which also provides the bias for transistor $Q_2$, while the output matching is established by resistor $R_6$.

A reference 50-Ω amplifier, utilizing two cascaded CE stages with different operating points was built to perform FID signal measurements at 3 MHz. The first stage provides a low NF, while the second stage delivers a high gain. The detailed bill of materials and PCB render are provided in Supplementary Figure 2.

**2.2 Experimental study**

To validate the simulation results, prototypes of the developed LNA were assembled. The PCBs were manufactured using a standard FR-4 substrate (1.5 mm thick, double-sided with 35-µm copper metallization). A list of component values and the main view of the LNA assembly board are available in Supplementary Figure 1. The TR1300/1 vector network analyzer – VNA (Copper Mountain Tech., USA, IN, Indianapolis) was used to evaluate the impedance characteristics and the $P_{1dB}$ compression point. NF measurements were performed in an anechoic chamber using an N8973A noise figure analyzer – NFA (Agilent-Keysight, USA, CA, Santa-Rosa) and a 16603DA noise source (Ceyear, Qingdao, China), while the LNA was fed by a low-noise HMP2030 power supply (Hameg, Frankfurt/Main, Germany). Since the available NFA is limited to a minimum frequency of 10 MHz, the NF at 3 MHz was characterized using a tinySA ULTRA spectrum analyzer (tinySA.org, Eindhoven, Netherlands), which has a built-in noise-figure meter covering frequencies below 10 MHz.

To evaluate the decoupling performance of the proposed LII preamplifiers, a simple loop coil imitating an element of an MRI coil array was fabricated on an FR-4 substrate. The inner diameter of the loop was 100 mm with a 4-mm conductor width (copper 25 µm). Circuit

No. 1 from (17) was implemented as the matching network, and the final connection diagram is shown in Figure 2. The circuit components and their values are presented in Figure 3a, while the setup for the isolation measurement is shown in Figure 3b. It consists of a dual-loop probe connected to the VNA and the loop antenna element, to which a 50-Ω load, a short circuit, and the LII LNA are connected in turn. To demonstrate the decoupling capabilities of the LNA $|S_{21}|$ parameter between the dual-loop probes was measured for both the LII and 50-Ohm amplifiers connected to the coil input.

To evaluate the performance of the LII LNA at 21.2 MHz, MR images of a rectangular aqueous phantom ($\varepsilon = 80$, $\sigma \approx 0.01$ S/m) were obtained using a single circular loop in Rx-only mode connected to a TOMIKON 0.5T MRI scanner (Bruker, Billerica, MA, USA). The experimental setup is shown in Figure 4. The MR images of the phantom were acquired using a gradient-echo (GRE) pulse sequence (TR/TE = 200/15 ms, matrix size = 64x64, FOV = 300x300 mm$^2$, slice thickness = 10 mm) for three configurations: without an external LNA, with the proposed LII preamplifier prototype, and with the coil connected directly to the Bruker RF unit with built-in LNA (the standard setup). During imaging, the proposed LII LNA was located a few meters away from the MRI scanner bore, near the receiver block.

At 3 MHz LII LNA was tested using a custom-built ultra-low-field (ULF) MR spectrometer, where FID signals were obtained using both the LII and the 50-Ohm LNAs. For transmission, a large solenoidal coil was used (number of turns N = 5, 0.8-mm wire diameter, 220-mm coil diameter). For reception, a single Rx-only identical to the one used for the preamplifier decoupling measurements was used. The signal source was a 2-L cylindrical phantom containing 1.25 g of $NiSO_4 \cdot 6H_2O$ and 5 g of NaCl per 1000 g of distilled $H_2O$. Using the acquired FID data, the SNR was calculated as the ratio of the peak signal intensity to the standard deviation of the noise for both the LII and 50-Ω amplifiers.

## 3 RESULTS

### 3.1 Numerical simulations

The input impedance of a CB circuit strongly depends on the transistor bias. For a BJT, an increase in the collector current leads to a decrease in the input impedance. The bias conditions can accordingly be set to ensure an input impedance of 3 Ω. Specifically, for the 3 MHz LNA, the collector current $I_c$ was set to 12.8 mA with a base-emitter voltage $V_{be}$ of 830 mV.

At 21.2 MHz, increasing the $I_c$ currents above 14 mA reduces the input impedance below 3 Ω; however, this further increase elevates the NF beyond 0.9 dB. Conversely, when $I_c$ is reduced below 10 mA, the input impedance increases to 3.5 Ω, and the NF reaches 0.8 dB. As a trade-off, in simulations, $I_c$ current was selected as 14 mA ($V_{be}$ = 819.4 mV) to achieve an input impedance of 3 Ω with an NF of 0.94 dB.

## 3.2 Experimental studies

### *3.2.1 On-bench measurements*

Figure 5a shows S-parameter curves for a 3-MHz LNA, illustrating good agreement between simulation and experimental results. In this case, the power gain $|S_{21}|^2$ is 43.2 dB while the input impedance reaches a maximum of 3 Ω at 3 MHz due to the matching circuit (Figure 5b). $S_{22}$ value is below -30 dB in the 2 MHz band. The measured compression point $P_{out,1dB}$ for the 3-MHz LNA is 8 dBm. An NF value estimated in the experiment matches the simulation data, taking into account the measurement errors, and is equal to 1.3 ± 0.2dB at 3 MHz (Figure 6a). In the dual-loop probe setup using the LII LNA, the preamplifier decoupling reaches -16 dB with a 50 Ω load (Figure 7a).

For the 21.2 MHz LNA, as shown in Figure 5d, the input impedance is predominantly real and remains equal over a 2 MHz range. After adjusting the bias and matching circuit, the gain of the 21.2 MHz LNA reaches 45.5 dB. Excellent output impedance matching is achieved with a 51 Ω resistor, and $S_{22}$ is below -13 dB over a 2 MHz band (Figure 5c). The $P_{out,1dB}$ of this LNA is 5.8 dBm in simulation and 5 dBm in experiment. The measured NF matches the simulation data across a wide frequency band and equals 0.87±0.05 dB at 21.2 MHz (Figure 6b). An experiment verifying the preamplifier decoupling capabilities of the 21.2 MHz LII LNA showed a decrease of 18 dB in the dual-loop probe $|S_{21}|$ relative to a 50 Ω load. For comparison, the lowest achievable $|S_{21}|$ with a short-circuit unit is -39 dB (Figure 7b).

Based on the 3 MHz results of this work, a further study was conducted on an LNA configuration consisting solely of a CB stage. The motivation was to reduce the NF, which exceeded the simulated values by 0.3 dB at 3 MHz. Figure 8b shows the input impedance, which agrees well with the simulation. A gain of 16.4 dB was measured for the prototype (Figure 8a). This CB LNA was matched at the output with the $|S_{22}|$ reaching -30 dB. Given this return loss, the measured NF was as low as 0.8 dB (Figure 8c). We therefore conclude that the increased noise in the "inverted cascode" version is due to a mismatch between the first and second stages. Indeed, the CE circuit has an input impedance of approximately 123 Ω, which corresponds to a return loss of -7.5 dB. To eliminate this drawback, a three-element matching circuit can be employed.

### *3.2.2 0.5T MRI experiment*

Further evaluations included MR imaging using a single loop to demonstrate the SNR obtained with the proposed LNA and to compare it with that of the reference preamplifier in a Bruker MRI receive block. Figure 9 shows water-phantom MR images obtained with a GRE sequence. Without an LNA (Figure 4a), the SNR is low as expected (SNR = 7), whereas with the reference built-in Bruker preamplifier, it improves by a factor of 40 (SNR = 284). Meanwhile, with the proposed LNA, a 55-fold signal amplification is achieved, resulting in an SNR of 386. To illustrate the difference, SNR profiles are plotted along the phantom's central transverse plane (Figure 9d).

### *3.2.2 0.07T FID experiments*

The free induction decay signals obtained with our LF-system are shown in Figure 10. There are no visible differences between the LII and 50 the Ohm amplifiers in either the time or frequency domain, indicating that the proposed design operates correctly at these frequencies.

## DISCUSSION

In this study, the design and performance of a CB bipolar transistor circuit as the first stage of a low-noise LF MRI preamplifier are demonstrated. A benefit of the proposed design is the combination of a low NF and LII in the HF band. Simulations and experimental studies showed that a CB circuit with the low-cost BFR193 transistor achieves an NF of 1.3 dB and a $Z_{in}$ of 3 Ω at 3 MHz, and an NF of 0.9 dB with the identical $Z_{in}$ at 21.2 MHz. This appears to be a reasonable trade-off to achieve an optimal LNA configuration.

In the study, extreme cases were considered: exceptionally low NF and exceptionally low input impedance. It was shown that for the minimum NF value, the collector current should be reduced. If the value is as low as 10 mA, the NF is below 0.8 dB (Figure S3). However, the input impedance is close to 5 Ω, which is insufficient for proper preamplifier decoupling. If we aim for a $Z_{in}$ reduction to 2 Ω, NF increases by over 2 dB. The circuit's current consumption exceeds 50 mA in this case due to the low input impedance. It should also be noted that the NF and the $Z_{in}$ can be reduced by connecting several transistors in parallel (18). This technique has been well-established and commonly used at low frequencies, and can also be used for further noise and input impedance reduction in our case. The final option for circuit optimization is to use another NPN transistor. Based on our experience, the Infineon BFR740 bipolar transistor outperforms the BFR193 in noise characteristics, but is less linear and more expensive. However, in practice extreme NF values are not required, allowing for reduction of $Z_{in.}$ This is because the self-NF (19) of the antenna element (or coil) exceeds 3 dB at low frequencies, thereby relaxing the NF requirements for the preamplifier.

## CONCLUSION

The results of this work demonstrate the operation of a low-noise amplifier composed of cost-effective BJTs. with the circuit features a common-base input stage that provides low input impedance, which is crucial for preamplifier decoupling methods in phased arrays for low-field MRI. The proposed LNA design can be adapted to any field strength from 0.1 to 1T by changing the matching circuit elements. Simulation results were verified experimentally, including phantom MR imaging at 0.5T and FID measurements at 0.07T. The use of accessible components in the proposed LNA broadens its application scope in MRI systems in low-income regions. The PCB design used in this work is an open-source project and is publicly available in the GitHub repository.

## ACKNOWLEDGMENTS

This work was supported by state assignment No. FSER-2025-0018 within the framework of the national project "Science and Universities"

## DATA AVAILABILITY STATEMENT

The PCB design files that support the findings of this study are openly available in a GitHub repository at https://github.com/Aleksei-Nasonov/LII_LNA.

**FIGURES**

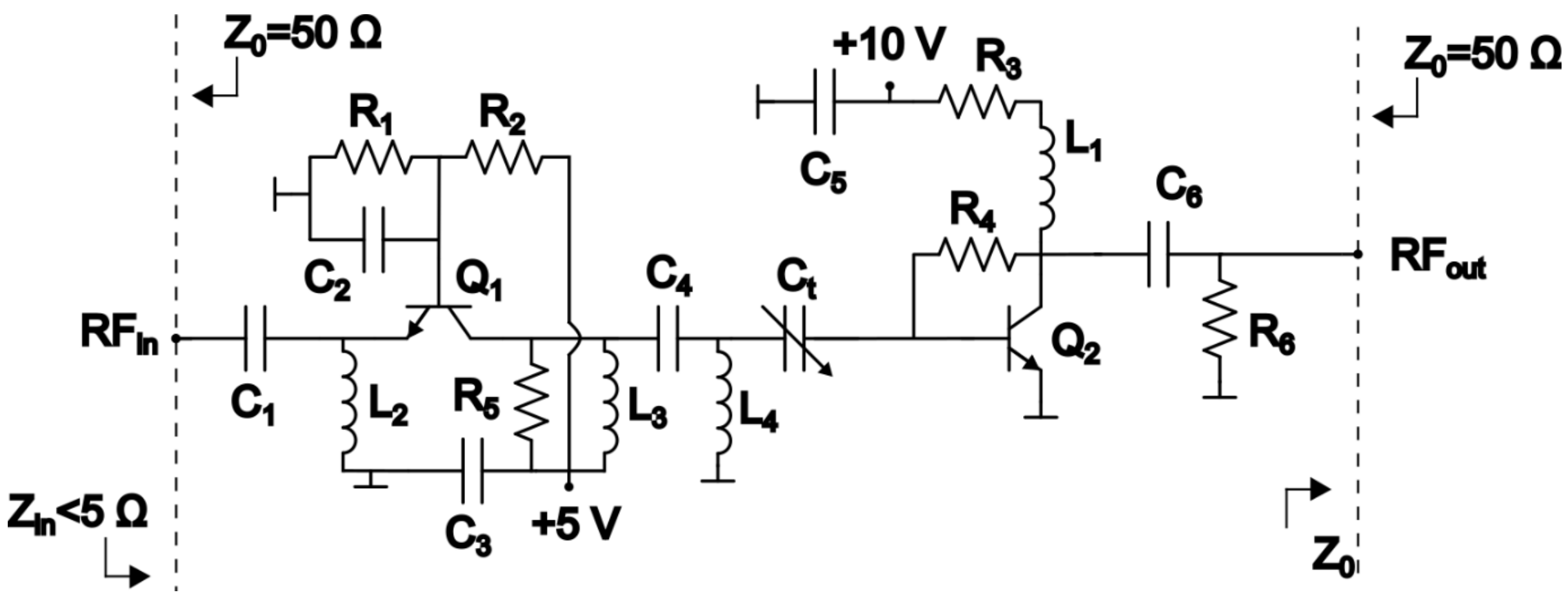

Figure 1: A two-stage LNA circuit with low input impedance, where Q1 is connected in a common base configuration and Q2 in a common emitter configuration

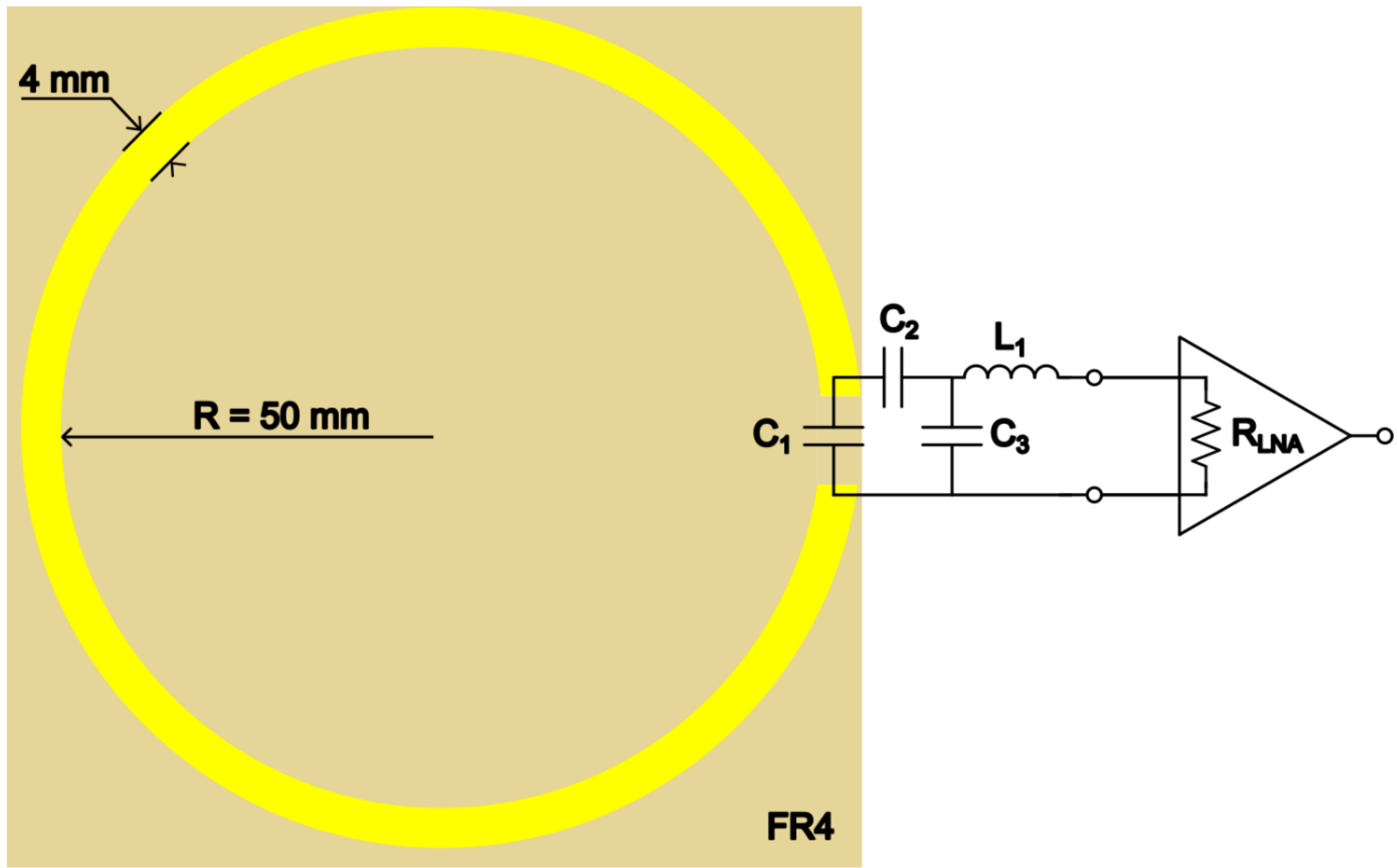

Figure 2: Test loop with a matching circuit for evaluating preamplifier decoupling at 21.2 MHz with low input impedance ($R_{LNA}$=3 Ω) LNA

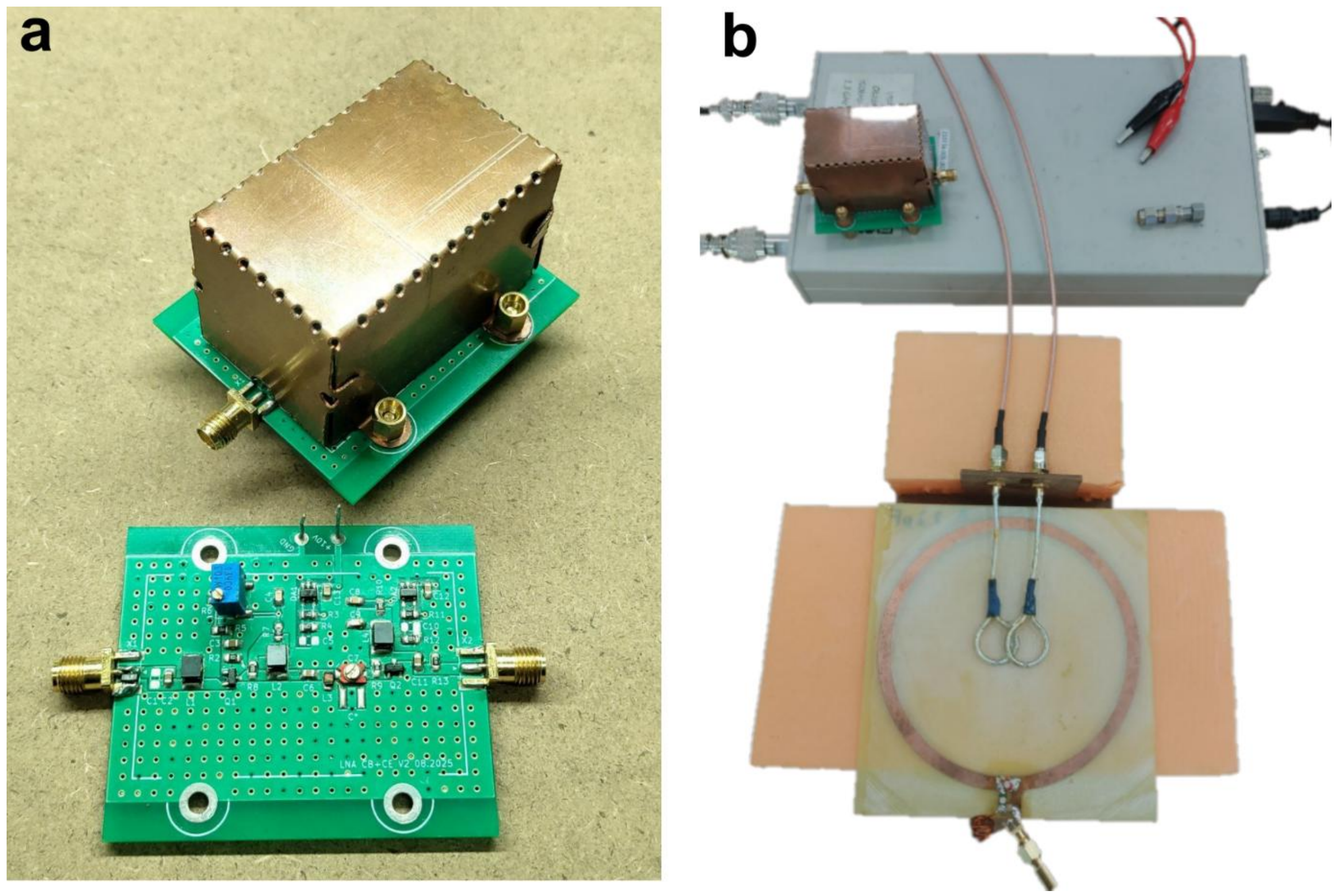

Figure 3: Prototypes of LNAs with and without copper shield (a); a setup for measuring preamplifier decoupling, consisting of a vector network analyzer, double probe and loop coil (b)

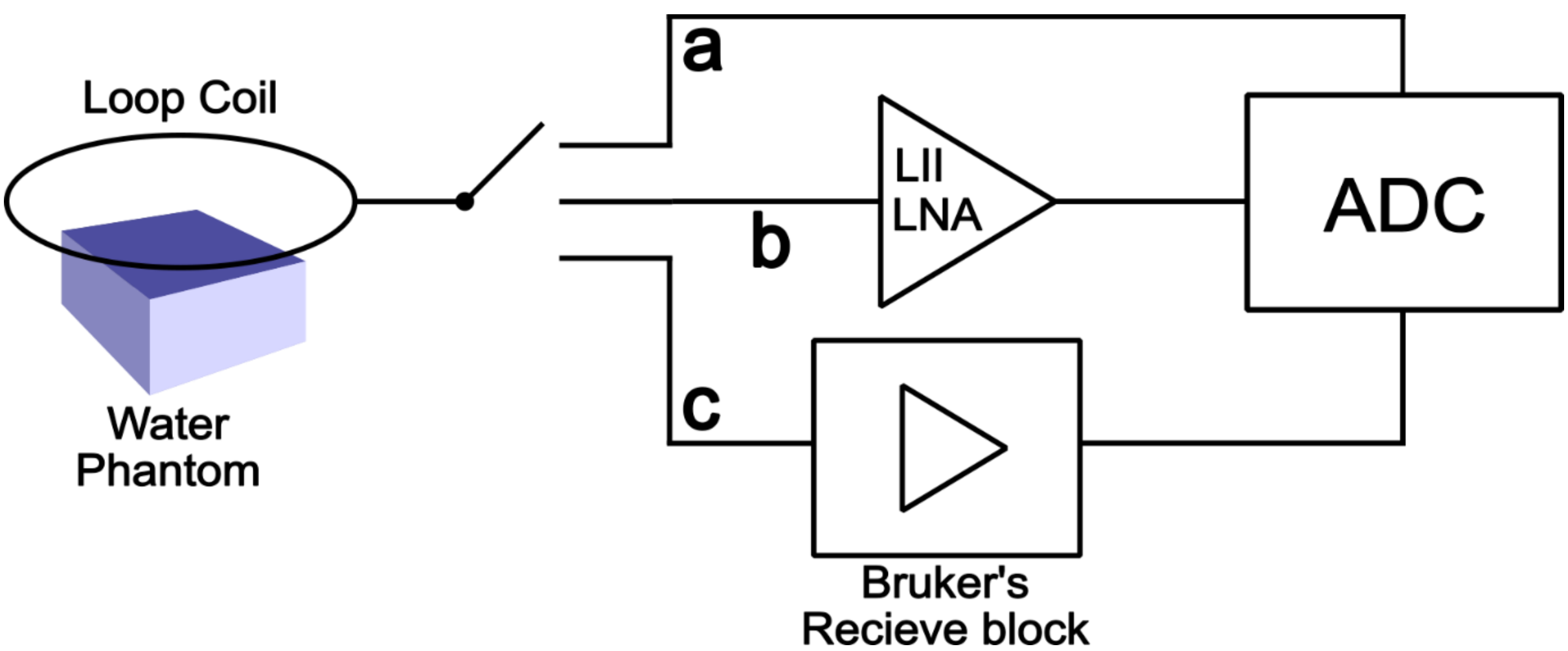

Figure 4: Design of the experimental setup for obtaining images on a 0.5T MRI scanner with the RF coil connected: directly to the ADC, bypassing the receiver unit (a); through a low input impedance LNA (b); via a standard connection through the scanner's receiver unit (c).

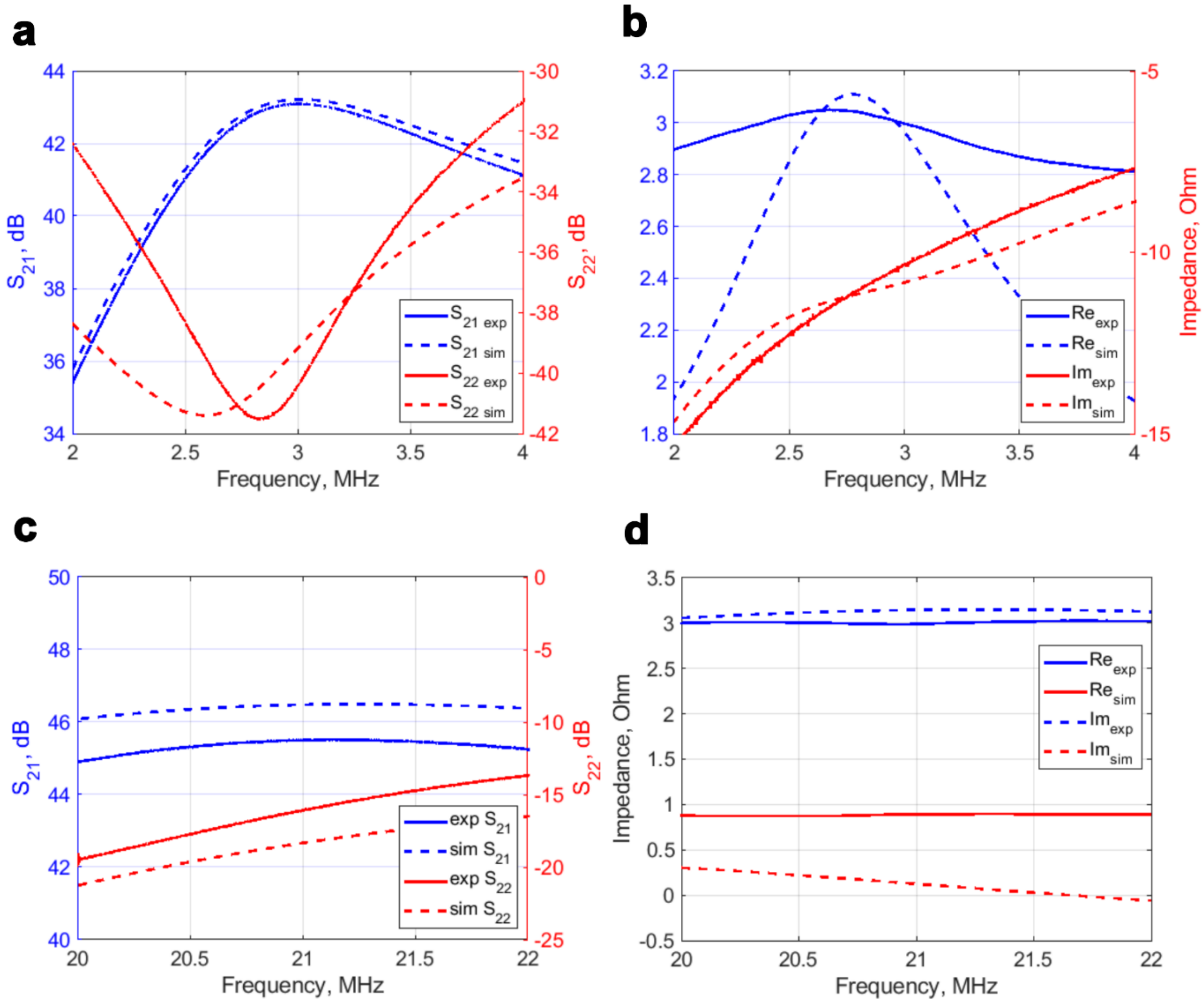

Figure 5: Simulated and measured $S_{21}$ and $S_{22}$ parameters (a) and input impedance (b) spectra for the 3 MHz LNA version; the same data is shown for the 21.2 MHz LNA version (c) and (d).

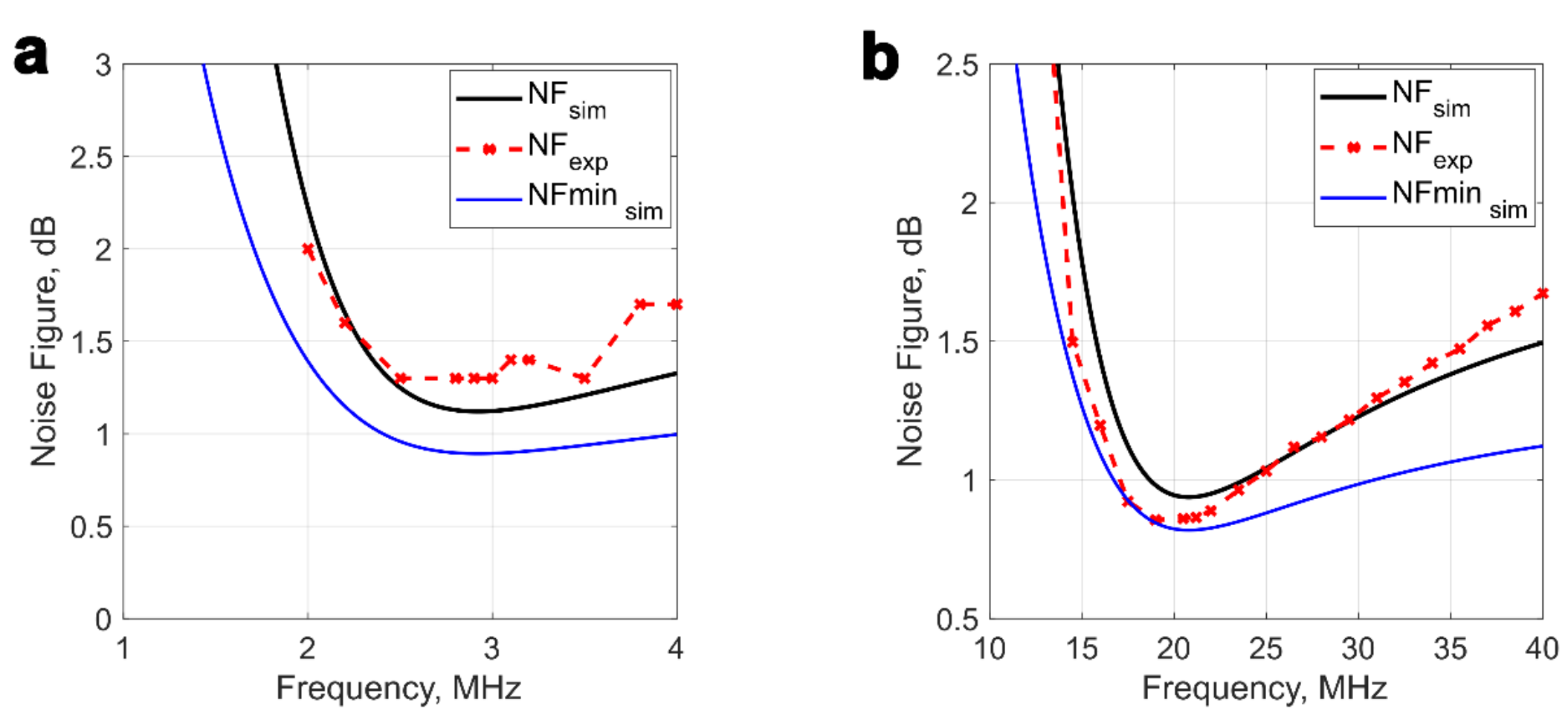

Figure 6: The noise figure spectra obtained from simulations and experiments for the 3 MHz LNA (a) and the 21.2 MHz LNA(b)

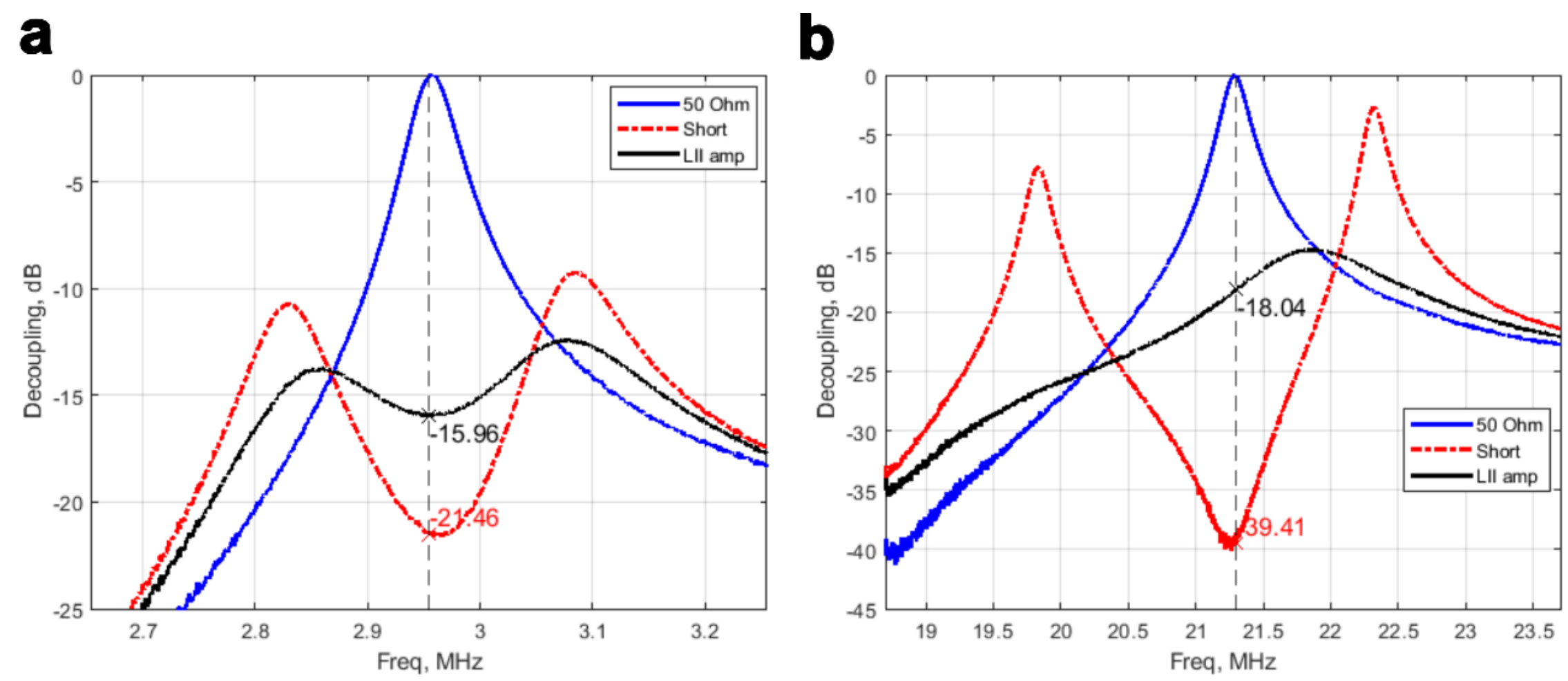

Figure 7: Comparison of experimentally obtained preamplifier decoupling for test loops at 3 MHz (b) and 21.2 MHz (e), versus cases with coils terminated with a calibration short and a reference 50 Ohm load, all normalized to the 50-Ohm load.

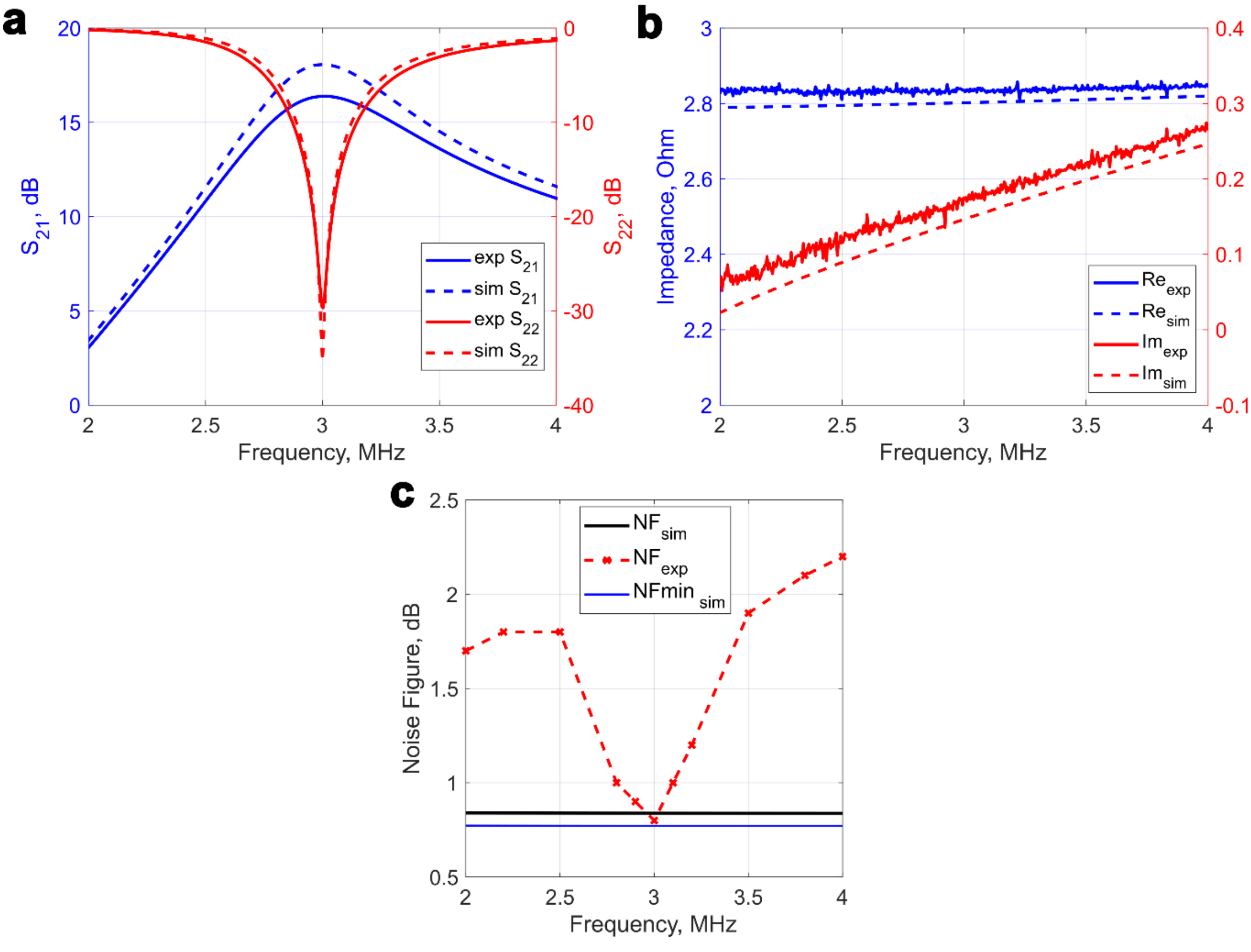

Figure 8: Simulated and measured $S_{21}$ and $S_{22}$ parameters (a) and input impedance (b) spectra for the 3 MHz common-base-only LNA; simulated and measured noise figure spectra (c).

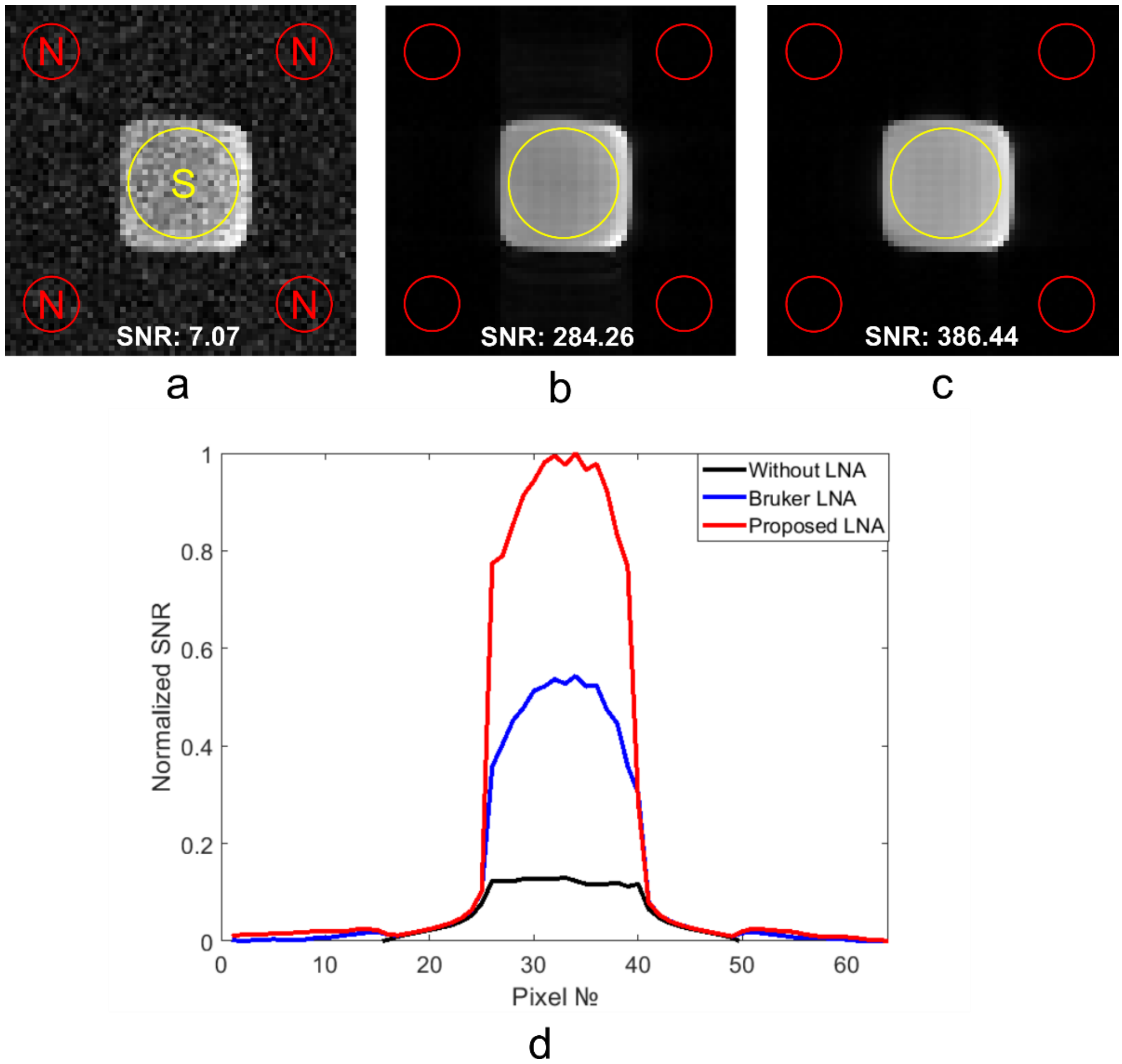

Figure 9: MR images of the water phantom obtained using a 0.5T Bruker MRI scanner and a spin-echo sequence with the SNR values, which were calculated as the ratio of the mean signal (S) to the noise, where the noise was defined as the standard deviation of the background noise (SD) measured in four corner regions of the image. Without LNA (a), Bruker's LNA (b), proposed LNA (c); Normalized SNR profiles through the center of the water phantom in the transverse direction, obtained from the images (d)

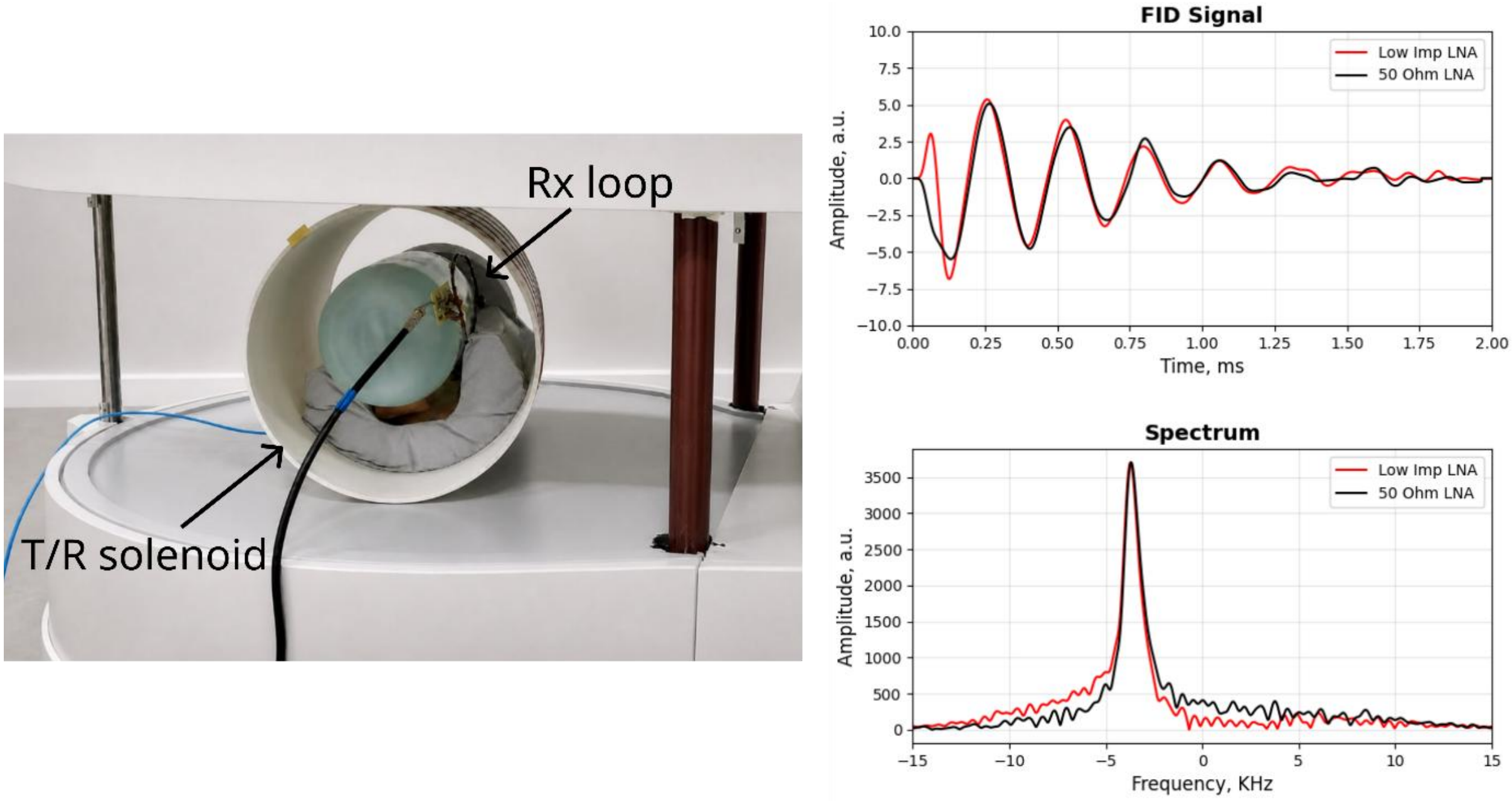

Figure 10: FID signal and its spectrum acquired with the developed low-field MRI system (0.07T permanent magnet), using a low-impedance LNA compared to a reference 50 Ohm LNA.

**SUPPLEMENTARY MATERIAL**

|  | References | Value | Quantity |
|---|---|---|---|
| 1 | C9, C12, C13 | 100 nF | 3 |
| 2 | C1, C3, C6, C11 | 220 nF | 4 |
| 3 | C4, C8 | 10 uF (tantal) | 2 |
| 4 | C5, C10 | - | 2 |
| 5 | C2 | 10 pF | 1 |
| 6 | C7 | C Trim 5-20pF | 1 |
| 7 | C7* | C match | 1 |
| 8 | R3, R11 | 6040 Ohm | 2 |
| 9 | R2 | 1.8K | 1 |
| 10 | R4 | 18 k | 1 |
| 11 | R5 | 4.7 K | 1 |
| 12 | R6 | 10 K | 1 |
| 13 | R7 | 47 Ohm | 1 |
| 14 | R8 | 3.9 K | 1 |
| 15 | R9 | 30 K | 1 |
| 16 | R10 | 51 Ohm | 1 |
| 17 | R12 | 43 Ohm | 1 |
| 18 | R13 | 62 Ohm | 1 |
| 19 | L1, L2, L4 | 100 uH | 3 |
| 20 | L3 | L match | 1 |
| 21 | D1, D2 | BAV99 | 2 |
| 22 | X1, X2 | SMA conn. | 2 |
| 23 | DA1, DA2 | MIC5205YM5-TR | 2 |
| 24 | Q1, Q2 | BFR193E | 2 |

Figure S1: Bill of materials and PCB render for the LNA with low input impedance.

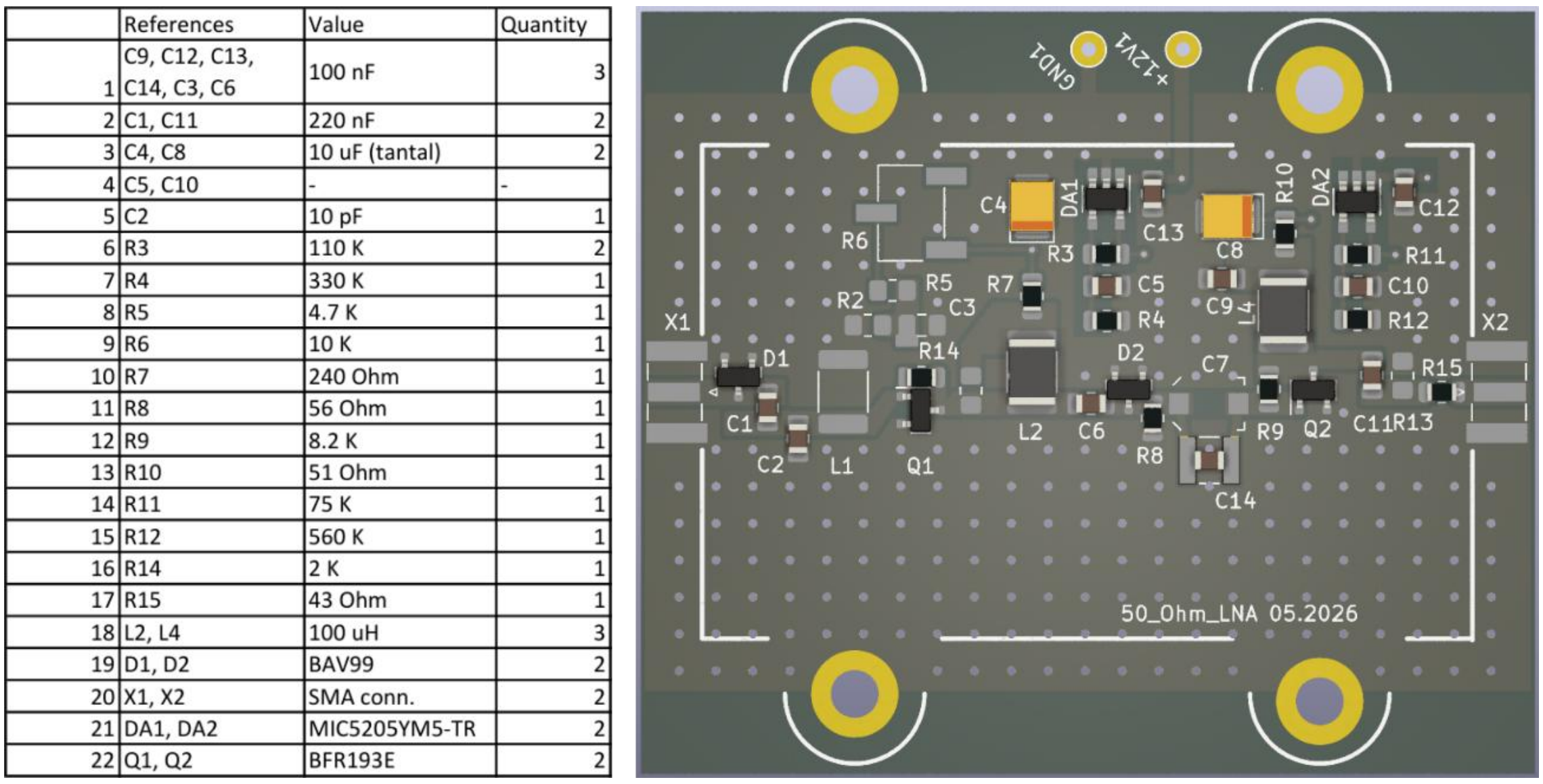

|  | References | Value | Quantity |
|---|---|---|---|
| 1 | C9, C12, C13, C14, C3, C6 | 100 nF | 3 |
| 2 | C1, C11 | 220 nF | 2 |
| 3 | C4, C8 | 10 uF (tantal) | 2 |
| 4 | C5, C10 | - | - |
| 5 | C2 | 10 pF | 1 |
| 6 | R3 | 110 K | 2 |
| 7 | R4 | 330 K | 1 |
| 8 | R5 | 4.7 K | 1 |
| 9 | R6 | 10 K | 1 |
| 10 | R7 | 240 Ohm | 1 |
| 11 | R8 | 56 Ohm | 1 |
| 12 | R9 | 8.2 K | 1 |
| 13 | R10 | 51 Ohm | 1 |
| 14 | R11 | 75 K | 1 |
| 15 | R12 | 560 K | 1 |
| 16 | R14 | 2 K | 1 |
| 17 | R15 | 43 Ohm | 1 |
| 18 | L2, L4 | 100 uH | 3 |
| 19 | D1, D2 | BAV99 | 2 |
| 20 | X1, X2 | SMA conn. | 2 |
| 21 | DA1, DA2 | MIC5205YM5-TR | 2 |
| 22 | Q1, Q2 | BFR193E | 2 |

Figure S2: Bill of materials and PCB render for the 50 Ω reference version LNA.

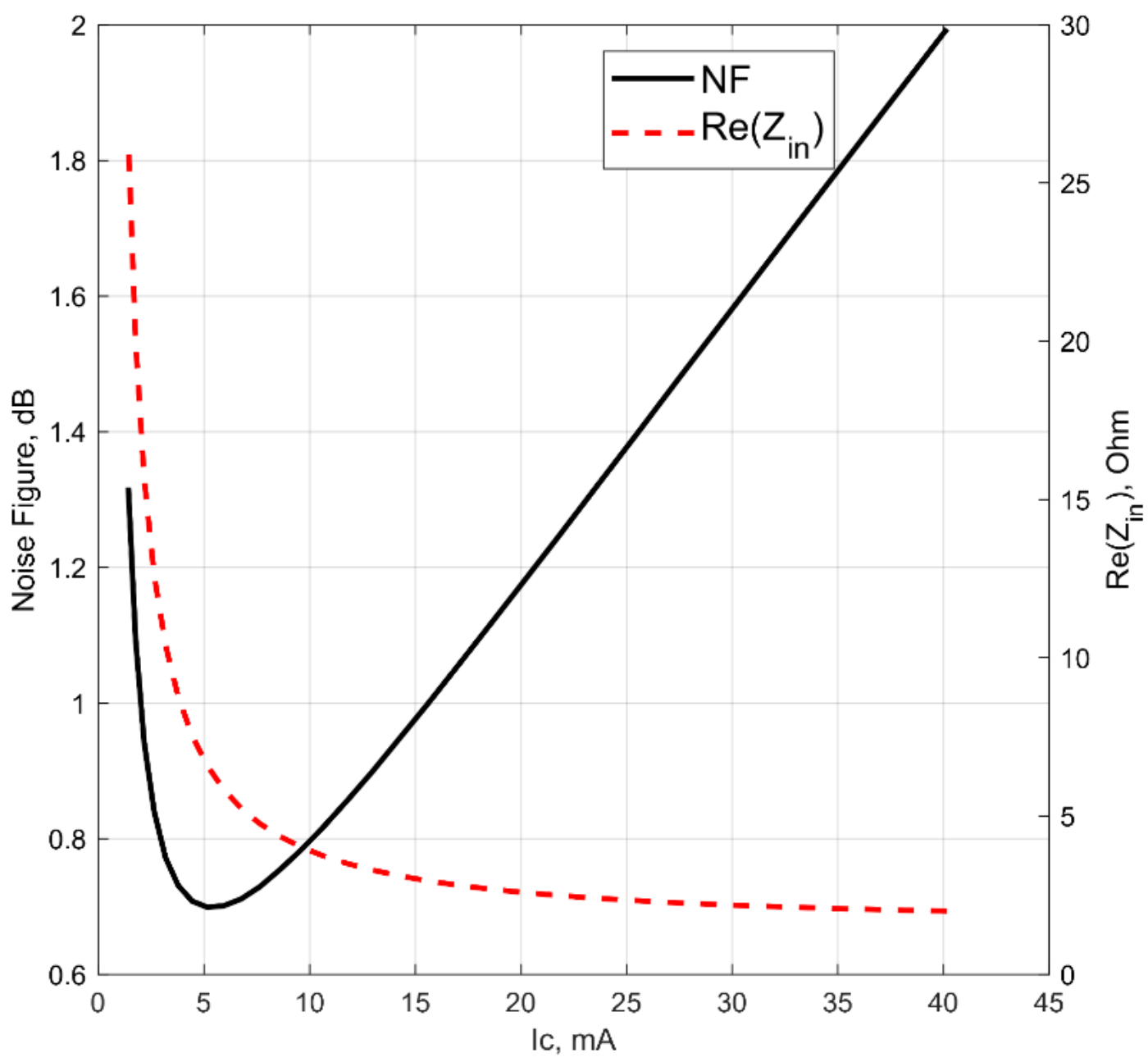

Figure S3: Dependence of the noise figure and real part of the input impedance $Z_{\mathrm{in}}$ of the "inverted cascode" LNA on the common-base collector current.